\documentstyle[11pt,epsf]{article}

\input psfig
\def\beq{\begin{eqnarray}}
\def\eeq{\end{eqnarray}}
\def\bea{\begin{eqnarray*}}
\def\eea{\end{eqnarray*}}

\def\centeron#1#2{{\setbox0=\hbox{#1}\setbox1=\hbox{#2}\ifdim
\wd1>\wd0\kern.5\wd1\kern-.5\wd0\fi
\copy0\kern-.5\wd0\kern-.5\wd1\copy1\ifdim\wd0>\wd1
\kern.5\wd0\kern-.5\wd1\fi}}
\def\ltap{\;\centeron{\raise.35ex\hbox{$<$}}{\lower.65ex\hbox{$\sim$}}\;}
\def\gtap{\;\centeron{\raise.35ex\hbox{$>$}}{\lower.65ex\hbox{$\sim$}}\;}

\def\singleandthirdspaced{\baselineskip=\normalbaselineskip\multiply
    \baselineskip by 130\divide\baselineskip by 100}
\def\singlespaced{\baselineskip=\normalbaselineskip}

\newcommand{\newc}{\newcommand}
\newc{\qbar}{{\overline q}}
\newc{\Kahler}{K\"ahler }
\newc{\deltaGS}{\delta_{\rm GS}}
\begin{document}
\begin{titlepage}
\begin{flushright}
{\large
hep-th/9810021 \\
SCIPP-98/32\\

}
\end{flushright}

\vskip 1.2cm

\begin{center}


{\LARGE\bf Tree Level Supergravity and the Matrix Model}

\vskip 1.4cm

{\large
Michael Dine, Robert Echols and Joshua P. Gray}
\\
\vskip 0.4cm
{\it Santa Cruz Institute for Particle Physics,
     Santa Cruz CA 95064  } \\

\vskip 4pt

\vskip 1.5cm

\begin{abstract} 
It has recently been shown that the Matrix model and
supergravity give the same predictions for three graviton
scattering.  This contradicts an earlier claim in the literature.
We explain the error in this earlier work, and go on to
show that certain terms in the $n$-graviton scattering
amplitude involving $v^{2n}$
are given correctly by the Matrix model.
The Matrix model also generates certain $v^6$ terms in four
graviton scattering at three loops,
which do not seem to have any counterparts in supergravity.
The connection of these results with nonrenormalization theorems
is discussed.

\end{abstract}

\end{center}

\vskip 1.0 cm

\end{titlepage}
\setcounter{footnote}{0}
\setcounter{page}{2}
\setcounter{section}{0}
\setcounter{subsection}{0}
\setcounter{subsubsection}{0}

\singleandthirdspaced

\section{Introduction}

One of the important pieces of support for the original
Matrix model conjecture was that it successfully reproduced
graviton-graviton scattering in supergravity\cite{bfss,reviews}.  Subsequently,
the model has been shown to reproduce the full helicity-dependent
amplitude\cite{waldron}.
As suggested in \cite{bfss}, this agreement can
be understood in terms of non-renormalization 
theorems\cite{beckers,ds,sethi,echols}.  That this agreement holds 
for finite as well as infinite
$N$ is an important component of the stronger DLCQ conjecture\cite{dlcq}.
Seiberg\cite{seiberg} and Sen\cite{sen}
have shown that this conjecture follows from well-established
duality relations.  However,
their argument does not necessarily imply that tree level supergravity
amplitudes should agree with the leading matrix model
computation.  Some issues related
to these proofs have been discussed in \cite{ph,banksreview}.
A puzzle raised by these arguments in the case of
propagation in curved backgrounds has been discussed in
\cite{ooguri}.

In \cite{arvind}, a technique was suggested
to extract certain terms in the supergravity and
matrix theory $S$-matrices for more than
two gravitons.  The idea was to consider a hierarchy
of distance scales (impact parameters).  In the case of three
gravitons, for example, one takes one of the gravitons
to be far from the other two.  In the matrix model,
where the graviton separation translates into
frequencies of (approximate) harmonic oscillators,
one can then analyze
the problem
by first integrating out the most massive modes, and then
study the interactions of the remaining light degrees of
freedom.  In momentum space, the hierarchy of impact
parameters translates into a hierarchy of momentum transfers.
This yields an appreciable simplification of the
supergravity calculation as well.

Using these methods,
it was argued in \cite{arvind} that there was a discrepancy
between the computation of  three graviton scattering in the
matrix model and in tree level supergravity.
Calling the large distance $R$ and the smaller distance $r$,
and denoting the velocity of the far-away graviton by $v_3$,
the supergravity $S$-matrix contains a term (after Fourier transform):
\beq
v_3^4 v_{12}^2 \over r^7 R^7.
\eeq
However, it is straightforward to see, by power counting arguments
(which we will review in the next section) that no such term
can be generated in the matrix model {\it effective action}.
The authors of \cite{arvind} then went on to argue that this
term could not appear in the Matrix model {\it S-matrix}.

Subsequently, however, Taylor and Van Raamsdonk\cite{wati}\
pointed out, using simple symmetry considerations,
that if one writes an effective
action for gravitons in supergravity, this
action {\it cannot} contain such terms.  (Other
criticisms appeared in \cite{ferretti}.)
Shortly afterwards, Okawa and Yoneya\cite{yoneya}\ computed the
effective action on both the matrix model and supergravity sides,
and showed that there is complete agreement.  A related computation
appeared in \cite{wilkins}.  Other calculations have also
been reported recently showing impressive agreement
between the matrix model and supergravity\cite{danielsson}.

It is clear from these remarks that the difficulty in \cite{arvind}\
lies in extracting the Matrix model $S$-matrix from the effective action.
In the next section of this
note, we show how the ``missing term" is generated
in the $S$-matrix of the matrix model.  In order to do this
using the effective action approach, it is necessary to
resolve certain operator-ordering questions\footnote{The authors
of \cite{arvind} had convinced themselves that there was
no choice of operator ordering which generated the missing
term.  This was their basic error.}.  To deal
with these issues the most efficient approach is the
path integral.  In section 2, we review first the problem
of computing the $S$-matrix from the path integral
by studying small fluctuations about classical trajectories.
Once this is done, the isolation of the ``missing term'' is
not difficult.

Despite the error
in the analysis of \cite{arvind},
the method proposed there yields a considerable simplification
in the calculation of the effective action.  Indeed, it is
possible to calculate certain terms in just a few lines.   On the
supergravity side, there are also significant simplifications which
occur in this limit.  One might hope,
then, to extract general lessons from this approach.  For example,
one can easily  compare certain tensor structures in $n$-graviton
scattering, and perhaps try to understand whether (and why)
there is agreement.  One can also
try to examine, as in \cite{echols}\, the role of non-renormalization
theorems.

In section 3, then, we go on to compare
certain other terms in three graviton
scattering, some of which were not explicitly
studied in \cite{yoneya}.  These calculations can be performed extremely
easily using the methods proposed in \cite{arvind}, on both the matrix
model and supergravity sides, and are shown to agree.

Armed with this success, we consider
in section 4 scattering of more than three gravitons, and
scattering when more dimensions are compactified.
Some of the terms in the four graviton scattering amplitude
can readily be computed, and compared on both sides.
We find agreement of certain terms involving
eight powers of velocity.  We also find certain terms of order
$v^{2n}$ in
$n$-graviton scattering, for arbitrary $n$, agree.  On the other
hand, the matrix model at three loops generates
terms of order $v^6$ in four graviton scattering.
These do not have the correct scaling with $N$ to generate
a Lorentz invariant expression, and it is difficult to see
how they can be cancelled by other matrix model contributions
to the $S$-matrix.  These terms also indicate that there
are terms at order $v^6$ which are renormalized.

These observations raise a number of questions.  In particular, it is
not completely clear why the arguments of  \cite{seiberg}\  and \cite{sen}\
imply that the {\it classical}
supergravity amplitudes should agree with the matrix model result.
One might have thought that this should only hold in cases in which
there are non-renormalization theorems\cite{banksreview}.
These results indicate that already at the level
of the four graviton amplitude, there are not
non-renormalization theorems, at least in the most naive sense.
They also suggest that at order $v^{2n}$, the $n-1$ loop
matrix model diagram reproduces the supergravity amplitude,
but that there are discrepancies at three loops and beyond
in terms with fewer powers of velocity.
We will make some remarks on these issues in the final section,
but will not provide a definite resolution.

\section{Computing the $S$-Matrix in the Matrix Model}

The matrix model is the dimensional reduction of ten
dimensional supersymmetric Yang-Mills theory.  The action
is
\beq
S= \int dt  [ { 1 \over g} {\rm tr}( D_t X^i D_t X^i)
+ {1 \over 2g}M^6 R_{11}^2 {\rm tr}([X^i,X^j][X^i,X^j]) +
\eeq
$$~~~~~~{1 \over g} {\rm tr} (i \theta^T D_t \theta +
M^3 R_{11} \theta^T \gamma^i[X^i,\theta])  ]
$$
where $R_{11}$ is the eleven dimensional radius,
$M$ is the eleven dimensional Planck mass and $g=2R_{11}$.
The $\theta$'s are the fermionic coordinates.

At small transverse velocity and small momentum
transfer (with zero $q^+$ exchange) it is a simple
matter to compute graviton-graviton scattering
in the matrix model.  One considers widely separated
gravitons, and integrates out the high frequency modes
of the matrix model. This yields, at one loop, an effective Lagrangian
for the remaining diagonal degrees of freedom which behaves as
\beq
{\cal L}_{eff}={15 \over 16} { v^4 \over r^7}+ {\rm fermionic~terms}.
\eeq
If this effective lagrangian is then treated in Born approximation,
one reproduces precisely the supergravity result for the $S$-matrix.

Ref.\ \cite{arvind}\ focused on the problem of multigraviton
scattering in the matrix model.
For three graviton scattering, it is necessary to compute
the terms of order $v^6$ at two loops in the matrix model
Hamiltonian.  In the three graviton case, there are two
relative coordinates and correspondingly two relative velocities.
The basic strategy of \cite{arvind}, which will also be the strategy
of this paper, was to consider the case where one of the relative
separations, say $x_{13}=x_1-x_3=R$, was much larger
than $x_{12}=r$.  In this limit, oscillators with frequency
of order $R$ can be integrated out first, yielding an effective
lagrangian for those with mass (frequency) of order $r$ (or zero).
This effective lagrangian is restricted by $SU(2)$ symmetry.
Finally, one can consider integrating out oscillators with
mass of order $r$.

In computing the $S$-matrix for three graviton scattering,
as discussed already in \cite{arvind}, it is necessary
not only to compute the terms of order $v^6$ in the effective
action, but also to consider terms in the scattering amplitude
which are of second order in the {\it one loop} ($v^4$) effective
action.  In other words,  working with the effective
action, it is necessary to
go to higher order in the Born series.
 
In \cite{arvind}, it was observed that terms of the form
\beq
v_3^4 v_{12}^2 \over R^7 r^7
\label{R7r7}
\eeq
cannot appear in the {\it effective action} of the matrix
model.  This is seen by simple power counting arguments.  The
$v_3$ factors can only arise from couplings to heavy
fields.  Integrating out the fields with mass of order $R$
at one loop, the leading terms involving
the light fields $x^a$ are\cite{echols,echogray} of the
form $v_3^4 x^a x^a / R^9$.  Moreover,
it was argued that the terms in (\ref{R7r7}) were not generated by the
higher order Born series referred to above.  This last point,
however, is incorrect, and is the source of the error.
In fact, it is easy to find the corresponding term in the matrix model
$S$-matrix.

\begin{figure}[htbp]
\centering
\centerline{\psfig{file=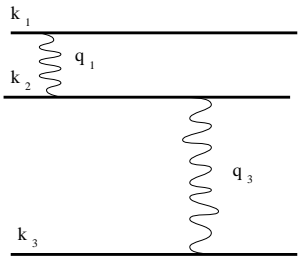,width=4cm,angle=-90}}
\caption{Ladder contribution to the supergravity amplitude.
Solid lines are the scattering gravitons.  Wiggly lines represent
virtual gravitons with zero longitudinal momentum.}
\label{sugraladder}
\end{figure}

\begin{figure}[htbp]
\centering
\centerline{\psfig{file=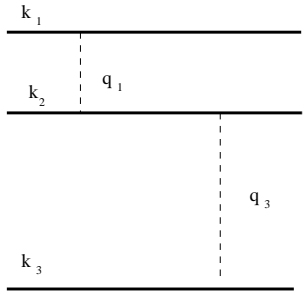,width=4cm,angle=-90}}
\caption{Corresponding contribution in the matrix model involving
iteration of the one loop Hamiltonian.
Dashed lines represent matrix elements of the
interaction potential.}
\label{secondborn}
\end{figure}

Consider the problem first from a Hamiltonian viewpoint.
We wish to compare the supergravity graph of fig.\ 1 with
the contribution of fig.\ 2 in old fashioned (time-ordered) perturbation
theory.  The second graph represents the iteration
of the one loop effective Hamiltonian to second order.
In momentum space, it has the correct
$1 \over q_3^2q_1^2$ behavior to reproduce
the $1\over R^7 r^7$ behavior of the missing supergravity
$S$-matrix term.  However, it has also an energy denominator,
and various factors of velocity.  
It is straightforward to check that this energy denominator is
proportional to
$1 \over {2k_2 \cdot q_1+q_1^2}$, the propagator appearing in the
covariant diagram of fig.\ 1.   To compare the diagrams in
more detail, one also needs
matrix elements of the type
$\langle \vec k_i + \vec q \vert H^{\prime}
\vert \vec k_i \rangle$
where $H^{\prime}$ is the one loop Hamiltonian.  The leading
term in powers of momentum transfer
is easily seen to reproduce the corresponding
term in the supergravity diagram.  In other words, if one ignores
the difference in the momenta of the particles in the initial,
final and intermediate states, one obtains exact agreement.
To see if higher order terms can cancel the
energy denominator and reproduce the
missing term,  it is necessary to keep at least terms linear in the
momentum transfer.  The problem is that it is not clear how
the momentum and $r$ factors are to be ordered in the
Hamiltonian.  Depending on what one assumes about this ordering,
one obtains quite different answers.

Of course, the full model has no such ordering problem.
It is only our desire to simplify the calculation using the effective
Hamiltonian that leads to this seeming ambiguity.  There is an
alternative approach, however, which leads to an unambiguous
answer, and where one can exploit the simplicity of the
one loop effective action.  This is to use the path integral.
As we will see, the path integral approach permits an
unambiguous resolution of the ordering problem.

\subsection{Path integral Computation of the S-Matrix}

Let us consider the problem of computing the $S$-matrix using the
path integral.  We will use an approach which is quite close to the
eikonal approximation (it is appropriate for small angle scattering)
which has been used in most analyses of
matrix model scattering.  It is helpful, first, to review some
aspects of potential scattering.  In particular, let us first see how
to recover the Born approximation by studying motion near
a classical trajectory.

A useful starting point is provided in \cite{waldron}.
In the path integral, it is most natural to compute the
quantity 
\beq
\langle \vec x_f \vert e^{-iHT} \vert \vec x_i \rangle = \int [dx] e^{iS}.
\eeq

To compute the $S$-matrix, one wants to take the
initial and final states to be plane waves, so one multiplies
by $e^{i \vec p_i \cdot \vec x_i}e^{-i \vec p_f \cdot \vec x_f}$
and integrates over $x_i$ and $x_f$.  For small angle scattering
in a weak, short-ranged
potential, one expects that the dominant trajectories are
those for free particles,
\beq
\vec x_o(t) = {\vec x_i + \vec x_f \over 2}+ \vec v t
\eeq
where $t$ runs from $-{T \over 2}$ to ${T \over 2}$,
and $\vec v= {\vec x_f- \vec x_i\over T}$.
It is convenient to change variables\cite{waldron}\ to
$\vec v$ and $\vec b$,
\beq
\vec b =  {\vec x_f + \vec x_i \over 2}.
\eeq
The complete expression for the amplitude is then
\beq
{\cal A}_{i \rightarrow f}= \int d^9 v \int d^9 b e^{i \vec b \cdot (\vec p_f-p_i)}
e^{i \vec v \cdot (\vec p_i + \vec p_f)} \int [d\vec x]e^{iS}.
\eeq
Now if we expand the classical action about this solution, writing
\beq
\vec x = \vec x_o + \delta \vec x
\eeq
(note $\delta \vec x$ includes both classical corrections
to the straight line path and quantum
parts)
we have a free piece
\beq
S_o = v^2 T/2.
\eeq
For large $T$, the $v$ integral can be done by stationary
phase, yielding
\beq
\vec v = {\vec p_f + \vec p_i \over 2}.
\eeq
We will see that this effectively provides the ordering prescription
we require for the matrix model problem.

For the case of potential scattering,
expand $e^{iS}$ in powers of $V$,
and replace the potential by its Fourier transform.
The leading semiclassical contribution to the amplitude
is then proportional to
\beq
\int d^9 be^{i \vec b \cdot (\vec p_f - \vec p_i)}
\int d^9q \int dt V(q) e^{i \vec q \cdot (\vec b + \vec v t)}.
\eeq
The $t$ integral gives a $\delta$-function for energy conservation,
while the $b$ integral sets $\vec q = \vec p_f - \vec p_i$.
This is precisely the Born approximation result.

Higher terms in the Born series can be worked out in a similar
fashion.  Time ordering the terms and replacing the
potential by its Fourier transform, the time integrals
{\it almost} give the expected
energy denominators.  The terms linear in momentum transfer
(involving $\vec v \cdot \vec q$) are given correctly, but
the $\vec q^2$ terms are not.  These terms must be generated
by the expansion of $V$ in powers of $\delta x$,
which generates additional powers of $\vec q$.  This problem,
which is essentially the problem of recoil discussed
in \cite{recoil},
will be analyzed in a separate publication \cite{rongli}.
Here we will work to leading order in $q$, and second order in
$V$.  

At second order in $V$, we need to consider an expression of
the form
\beq
\int dv \int db e^{i \vec v \cdot {\vec p_i + \vec p_f \over 2}}
e^{i \vec b \cdot (\vec p_f - \vec p_i)}
e^{-{v^2 T \over 2}} \int_{-T \over 2}^{T \over 2}dt_1
\int_{-T \over 2}^{T \over 2}dt_2 V(\vec x(t_1)) V(\vec x(t_2)).
\eeq
Time order the $t_1,t_2$ integrals, and Fourier transform each of the
factors of $V$.    The integral over $\vec v$ is again done by stationary
phase, and the resulting expression has the form:
\beq
{ 1 \over 2!}\int_{-T \over 2}^{T \over 2}dt_1
\int_{-T \over 2}^{t_1} dt_2
\int db \int dq_1 \int dq_2 {\cal V}(\vec q_1) {\cal V}(\vec q_2)
e^{i \cdot b (\vec p_f - \vec p_i) + i \vec q_1 \cdot (\vec b
 + \vec v t_1) 
+ i \vec q_2 \cdot ({\vec b + \vec v t_2})} 
\eeq
It is now straightforward to do the $t_i$, $\vec q_i$, and $\vec b$ integrals.
The integral over $t_2$ yields the energy denominator,
$1 \over \vec v \cdot \vec q_2$.  This differs from the exact energy
denominator by terms of order $q^2$.  The final integral over $t_1$
yields the overall energy conserving $\delta$-function.  Up to these terms
of order $q^2$, this is exactly the second order Born approximation
expression.

\subsection{The Ladder Graphs}

We are now in a position to compare the supergravity
and matrix model ladder graphs (see fig.\ 1 and 2).  On the supergravity
side, the calculation is completely standard, and proceeds
along the lines of \cite{arvind}.   As there,
we take the vertices from \cite{sanan}\footnote{To be
consistent with the authors \cite{yoneya,bbpt} who use
$\kappa^2 = 16 \pi^5$, the 3-vertex in \cite{sanan} needs to
be multiplied by $2$.}   
and require that the polarizations of the incoming
and outgoing gravitons be identical (as is true
to leading order in the inverse distance in the matrix
model).   
The $N_2 N_3{v_3^4 \over q_3^2}$ term
comes from the second vertex, and is precisely of the same
form as in graviton-graviton scattering. The
vertex on the first graviton line is
\beq
- k_{1 \sigma}(k_{1\gamma} - q_{1 \gamma})
- (k_{1 \sigma}-  q_{1 \sigma})k_{1\gamma} 
+ 2 k_{1 \sigma}k_{1 \gamma} + 2 (k_{1\sigma}-q_{1\sigma})(k_{1 \gamma}
-q_{1 \gamma}).
\eeq
>From the first vertex on the second graviton line, 
we get a similar expression,
replacing $k_1$ with $k_2$ and $q_1$ by
$-q_1$.  Multiplying these factors together, and including
the propagator, gives for the corresponding amplitude:
\beq
{\cal A}_1= (2\kappa)^4 (k_2 \cdot k_3)^2 (k_1 \cdot k_2)
{[(k_1 \cdot k_2)-(q_1 \cdot k_2) + {\cal O}(q_1^2)] \over
q_1^2 q_3^2 (2k_2 \cdot q_1 + q_1^2)}.
\eeq
Or
\beq
{\cal A}_1= {{\kappa}^4 \over 16} {N_1 N_2^2 N_3 v_3^4 v_{12}^2 [(v_{12}^2
-{2 \over N_1} \vec q_1 \cdot \vec v_{12}) + {\cal O}(q_1^2)]
 \over q_1^2 q_3^2(\vec q_1 \cdot \vec v_{12})},
\eeq
expressed in light cone variables with non-relativistic normalization.

Now we want to compare with the matrix model prediction,
fig.\ 2.  Recalling the averaging prescription, for the matrix elements
of the interaction Hamiltonian we have (dropping terms suppressed
by extra powers of $q_3$)
\beq
({15 \over 16})^2 N_1N_2^2 N_3 v_3^4 \left ( {\vec k_1 \over N_1}
-({\vec k_2 \over N_2}) -{1 \over 2} \vec q_1({1 \over N_1}
+ {1 \over N_2}) \right )^4
\eeq
\beq
~~~~= ({15 \over 16})^2 N_1 N_2^2 N_3 v_3^4 v_{12}^2 [( v_{12}^2
-{2 \over N_1} \vec q_1 \cdot \vec v_{12}) 
-{2 \over N_2}( \vec q_1 \cdot \vec v_{12})+ {\cal O}(q_1^2)].
\eeq
After Fourier transforming $r$ and $R$ (not shown above),  
the first term is exactly the term found on the supergravity side.
The second term cancels the energy denominator, yielding 
a contact term, 
\beq
-\kappa^4 N_1 N_2 N_3 {v_3^4 \over 8}{v_{12}^2 \over q_1^2 q_3^2}.
\eeq
Each of the four ladder diagrams yields an
identical contribution.  The sum
is precisely the ``missing'' term of \cite{arvind}.
At this level, there is no discrepancy between the DLCQ
prediction for the scattering amplitude
and supergravity.

\section{Additional contributions to three graviton scattering}

The method proposed in \cite{arvind}
should allow us to readily compute certain terms in the matrix
model scattering amplitude.  Consider, again, the case of three graviton
scattering, with
$R \gg r$.  It is very easy to obtain
the terms in the action involving
four powers of $v_3$ and two powers
of $v_{12}$ (fig.\ 3).  The point, again, is that
factors of $v_3$ can be obtained only from loops
involving fields with mass of order $R$,
and this costs powers of $R$; to obtain
the least suppression, one must attach
$v_{12}$ to the light fields.

\begin{figure}[htbp]
\centering
\centerline{\psfig{file=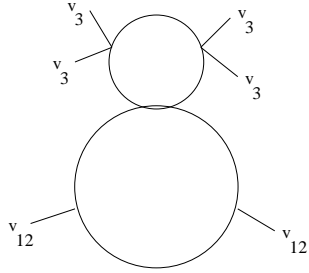,width=4cm,angle=-90}}
\caption{Matrix model contribution to three graviton amplitude.}
\label{matrixthree}
\end{figure}

\noindent
Integrating out fields with mass of order $R$
yields no terms independent of velocities or quadratic
in velocities; at quartic order in velocity, one has:
\beq
{\cal L}= {15 \over 16} ( {v_{13}^4\over \vert \vec x_{13} \vert^7}  
+  {v_{23}^4\over \vert \vec x_{23} \vert^7}).  
\eeq
For small $x_{12}$, one can
expand in powers of $x_{12}$.
The result can can be generalized to an $SU(2)$
invariant expression:
\beq
\delta{\cal L} = {15 \over 64}{v_3^4}
[((\vec x_1+\vec x_2)\cdot \nabla_R)^2 +
(\vec x^a \cdot \nabla_R)^2]{1 \over R^7}. 
\eeq  
Here $x_1+x_2$ is the center of mass of the $1-2$ system (combined
with the leading term, the expression is translationally
invariant).
The superscript $a$ is an $SU(2)$ index.  Contracting the 
$x^a$ factors, the leading (infrared divergent and finite) terms
cancels\cite{echols}.  The Euclidean propagator, up to
terms quadratic in
velocities is given by
\begin{equation}
\langle x^{+i} x^{-j} \rangle = {\delta^{ij} \over \omega^2+r^2} +  {4 (v^iv^j)
+ {\rm const}\ \delta^{ij} v^2 \over (\omega^2+r^2)^3}.
\label{prop}
\end{equation}
Substituting back in our expression above and performing
the frequency integral yields
\begin{equation}
N_1 N_2 N_3 {45 \over 64 R_{11}^5}{v_3^4 \over r^5} (\vec v_{12}\cdot \nabla)^2
{1 \over R^7}.
\label{matrix3}
\end{equation}
In deriving this expression it is necessary to keep track
of various factors of $2$.  One comes from the two real
massive fields in the loop (or equivalently, written in terms
of complex fields, from an extra $2$ which appears in the
vertex), the other from a factor of $g=2R_{11}$ for a 2-loop result.
It is easy to show that this is the only contribution with
this $r$ dependence and four factors of $v_3$.

\begin{figure}[htbp]
\centering
\centerline{\psfig{file=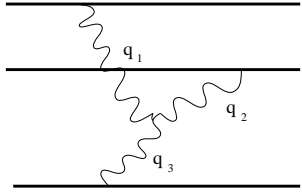,width=4cm,angle=-90}}
\caption{Contribution to three graviton amplitude involving
three graviton vertex.}
\label{sugrathree}
\end{figure}

Let's compare this with the supergravity amplitude.
There is only one diagram with the tensor structure of
(\ref{matrix3}); this arises from the diagram of fig. \ref{sugrathree}.
There are also several terms in individual diagrams of the form
$v_3^4  v_{12}^2 {1 \over R^8r^6}$, as well as
terms of order $1/R^9r^5$ with a different tensor structure
than the matrix model result.  We will shortly explain
that, at the level of the $S$-matrix, all of these terms
match, just as in the case of the leading $1/R^7r^7$ term.

Let us first consider the contribution to the supergravity
$S$-matrix of the form (\ref{matrix3}) above.
The relevant diagram is shown fig.\ 4.      
It is convenient to view $q_1$ and $q_3$ as independent,
so $q_2= -q_1-q_3$.  From \cite{sanan}, the necessary 
piece of the three graviton vertex is
\beq
2P_3(k_{1\sigma}k_{2\gamma}\eta_{\mu \nu} \eta_{\alpha\beta})
+2P_6(k_{1\sigma}k_{1\gamma}\eta_{\mu \nu} \eta_{\alpha\beta})
+4P_6(k_{1\nu}k_{2\gamma}\eta_{\beta\mu} \eta_{\alpha\sigma})
+4P_3(k_{1\nu}k_{2 \mu}\eta_{\beta \sigma} \eta_{\gamma\alpha}).
\eeq
It is then a straightforward exercise to evaluate the diagram.
Matters are considerably simplified by using kinematic
relations such as $k_1 \cdot q_1 = -{1 \over 2} q_1^2$,
$k_1 \cdot q_2 = {1 \over 2} q_1^2-k_1 \cdot q_3$,
etc., and dropping terms with the wrong $R$ dependence.
After only a few lines of algebra, this yields the covariant form
of the amplitude:
\beq
{16 \kappa^4 [(k_1 \cdot k_3)^2 (k_2 \cdot q_3)^2
+(k_2 \cdot k_3)^2 (k_1 \cdot q_3)^2-
2 (k_1 \cdot k_3) (k_2 \cdot k_3) (k_1 \cdot q_3)  (k_2 \cdot q_3)]} 
\over {q_1^4 q_3^2} .
\eeq
Changing to light cone variables with non-relativistic normalization
gives
\begin{equation}
{{\kappa^4 N_1 N_2 N_3} \over {2 R_{11}^3}} 
v_3^4 ((\vec v_1 - \vec v_2)\cdot \vec q_3)^2 {1 \over q_1^4 q_3^2}.
\label{sugra3}
\end{equation}
Then Fourier transforming gives precisely
the matrix model result (\ref{matrix3}).

There are several other kinematic structures which appear in
individual supergravity diagrams which do not arise in the
matrix model computation, and thus must be produced by iteration
of the one loop action.  The cancellation,
in fact, is closely related to the cancellation we have studied
of the leading term.  For example, there are terms from the diagram
of fig.\ 4 which behave as $v_3^4 v_{12}^2  \over R^8 r^6$.  To see how this
and other terms cancel, let us return to our earlier discussion.
There, we set $q_1=-q_2$.  However, we should be more careful,
and write $q_2 = -q_1 - q_3$.  Then from fig.\ 4 we have
a contribution
\beq
-{\kappa^4 \over 2}N_1 N_2 N_3 v_3^4 v_{12}^2
({q_1 \cdot q_2 + q_1 \cdot q_3 + q_2 \cdot q_3 \over q_1^2 q_2^2 q_3^2})
\eeq
(previously we kept only the first term and set
$q_1 = -q_2$).  We also have the supergravity term
involving the 4-vertex discussed in \cite{arvind} 
\beq
-{\kappa^4\over 2} N_1 N_2 N_3 v_3^4 v_{12}^2
({1 \over q_1^2q_3^2} + {1 \over q_2^2 q_3^2}).
\eeq
On the matrix model side, the higher order
Born terms yield
\beq
-{\kappa^4\over 4} N_1 N_2 N_3 v_3^4 v_{12}^2 
({1 \over q_1^2q_3^2} + {1 \over q_2^2 q_3^2}).
\eeq
As before, the leading terms match.  Expanding in powers of $q_3$,
it is not hard to check that the coefficients of $q_1 \cdot q_3$
and $(q_1 \cdot q_3)^2$ match as well.

\section{More Gravitons}

\subsection{n-Graviton Scattering}

Certain terms in the four and higher graviton scattering amplitude
are easily evaluated by these methods.  On the matrix model side, the
calculations are particularly simple.  One can, for example, consider
a generalization of the three graviton calculation above, indicated
in fig.\ 5.  At two loops, we saw that we generate in $SU(3)$
an effective coupling,
\beq
{45 \over 64} v_3^4 (\vec v_{12} \cdot \vec \nabla_R)^2{1  \over R^7 r^5}.
\eeq
We can generalize this to the case of $SU(4)$, with the
hierarchy $x_{4i} \gg x_{3\ell} \gg x_{21}$, where $i= 1,2,3$
and $\ell = 1,2$.  In other words, we again suppose that there
is a hierarchy of distance scales, with one particle very far from
the other three, and one of these three far from the remaining
two.  Again, we proceed by first integrating out the most massive
states, then the next most massive, and so on.  After the first two
integrations, we generate a term (among others)
\beq
{45 \over 64}v_4^4 (\vec v_3 \cdot \nabla_4)^2{1 \over \vert \vec x_4 \vert^7}
({1 \over \vert
\vec x_{31} \vert^5}
+{1 \over \vert \vec x_{32} \vert^5 }).
\eeq 
As before, expand this term in powers of the small
distances $x_1$, $x_2$, and generalize
to an $SU(2)$-invariant expression, yielding:
\beq
{45 \over 256}v_4^4 (\vec v_3 \cdot \nabla_4)^2 {1 \over \vert \vec 
x_4 \vert^7 }
(\vec x^a \cdot \nabla_3)^2 {1 \over \vert \vec x_3 \vert^5}. 
\label{sufour}
\eeq
Finally, the integration of $x_a$ yields various terms.  The piece
of  $\langle x_a^i x_a^j \rangle \propto v^i v^j$ yields
\begin{equation}
{135 \over 256}v_4^4 (\vec v_3 \cdot \vec \nabla_4)^2 {1 \over \vert
\vec x_4 \vert^7}
(\vec v_{12} \cdot \vec \nabla_3)^2 {1 \over \vert \vec x_3 \vert^5} 
{1 \over \vert \vec x_{12} \vert^5}.
\label{matrix4}
\end{equation}
Higher order terms corresponding to $n$-graviton scattering
generated in a similar fashion will be discussed below.

\begin{figure}[htbp]
\centering
\centerline{\psfig{file=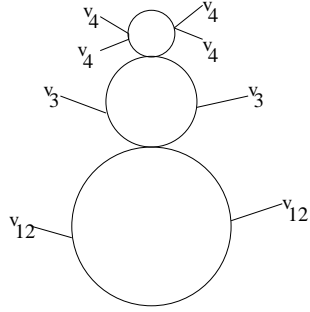,width=4cm,angle=-90}}
\caption{A matrix model diagram contributing to four graviton
scattering.}
\label{threeloopmatrixa}
\end{figure}

\begin{figure}[htbp]
\centering
\centerline{\psfig{file=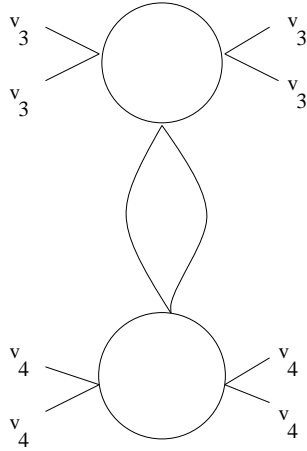,width=4cm,angle=-90}}
\caption{Another diagram which is easily computed.}
\label{threeloopmatrixb}
\end{figure}

Another term which is easily obtained is indicated in the diagram
in fig. 6.  This graph includes the interaction of the light $SU(2)$
fields from integrating out the fields with mass of order $x_4$ at
one loop, as well as those obtained by integrating out the fields
of mass $x_3$ at one loop.  The relevant interactions are
\beq
({15 \over 64})^2(v_4^4 v_3^4)
[ (\vec x^a \cdot \vec \nabla_4)^2 {1 \over \vert \vec x_4 \vert ^7} 
(\vec x^a \cdot \vec \nabla_3)^2 {1 \over \vert \vec x_3 \vert ^7} ]
\eeq
Now contracting the $x^a$ factors as in fig. 6 yields a term:
\begin{equation}
4({15 \over 64})^2 \left [
v_4^4v_3^4(\nabla_4^i \nabla_4^j) {1 \over \vert \vec x_4 \vert ^7} 
(\nabla_3^i \nabla_3^j) {1 \over \vert \vec x_3 \vert ^7} \right ]
{1 \over \vert \vec x_{12}
\vert^3}
\label{hard}
\end{equation}

\begin{figure}[htbp]
\centering
\centerline{\psfig{file=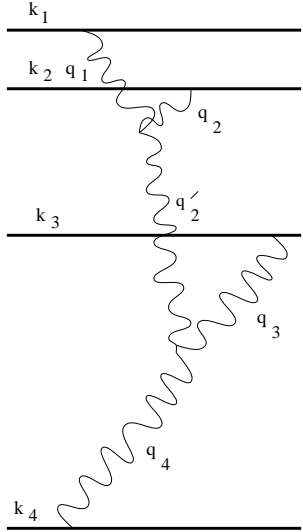,width=4cm,angle=-90}}
\caption{A supergravity contribution to four graviton scattering.}
\label{sugrafourgraviton}
\end{figure}

On the supergravity side, the required computations are somewhat more
complicated.  The easiest
to consider is the first term (\ref{matrix4}) above.
This term is generated by the diagram of fig. 7.
It is not difficult to find the particular tensor structures
which give the matrix model expression (\ref{matrix4}).  Focus first on the
terms involving $\vec v_3 \cdot \vec \nabla_4$.  These
must come from dotting $ k_3$ into $q_2^{\prime}$ or $q_4$. Calling
$q_2^{\prime}= -q_3-q_4$, the relevant term in the three graviton
vertex is ($\mu,\alpha$ are the polarization indices carried by
the graviton with momentum $q_2^{\prime}$)
\beq
2 \left [ P_3(q_{2 \sigma}^{\prime} q_{3 \gamma} \eta_{\mu \nu}
 \eta_{\alpha \beta})
+P_6(q_{2 \sigma}^{\prime} q_{2 \gamma}^{\prime}
\eta_{\mu \nu} \eta_{\alpha \beta})
+2P_6(q_{2 \nu}^{\prime} q_{3 \gamma} \eta_{\beta \mu} \eta_{\alpha \sigma})
+2P_3(q_{2 \nu}^{\prime} q_{3 \mu} \eta_{\beta \sigma} \eta_{\gamma\alpha})
\right ]
k_{4 \sigma} k_{4 \gamma} k_{3 \nu} k_{3 \beta} 
\eeq
Only a few permutations actually contribute, and one obtains simply
\beq
2(k_3 \cdot q_4)^2k_{4\mu }k_{4 \alpha}
\eeq
So the whole diagram collapses to $2(k_3 \cdot q_4)^2 \over q_3^4$
times the three graviton term we evaluated earlier.  The
result agrees completely with the matrix model computation (\ref{matrix4}).

Indeed, one can now go on to consider similar terms in $n$-graviton
scattering.   The supergravity graph indicated in fig. 8 can be evaluated
by iteration.  The coupling of the $n-1$ graviton is similar to that of
the third graviton in the $4$-graviton amplitude and can be treated
in an identical fashion.  The result then reduces to the $n-1$ graviton
computation.  So one obtains
\begin{equation}
{{15} \over {16}} ({{3} \over {4}})^{n-2}  
v_n^4 (\vec v_{n-1}\cdot \vec \nabla_n)^2{1 \over \vert \vec x_n \vert^7}
(\vec v_{n-2} \cdot \vec \nabla_{n-1})^2 {1 \over \vert \vec x_{n-1}\vert^5}
\dots (\vec v_3 \cdot \vec \nabla_4)^2{1 \over \vert \vec x_4 \vert^5}
(\vec v_{12} \cdot \vec \nabla_3)^2{1 \over \vert \vec x_3 \vert^5}
{1 \over \vert \vec x_{12} \vert^5}
\label{n-grav}
\end{equation}

The corresponding term in the matrix model effective action is also
obtained by iteration.  It is easy to generalize the calculation of
fig.\ 5 to the case above.  Repeating
our earlier computations gives precisely the 
result of eqn.\ (\ref{n-grav}) above.
 
The computation of the part of the supergravity amplitude
corresponding to eqn.\ (\ref{hard}) is more complicated.  This
term is generated by the sum of  several diagrams.  We will not attempt
a detailed comparison here, leaving this, as well as certain other terms, for
future work.

\begin{figure}[htbp]
\centering
\centerline{\psfig{file=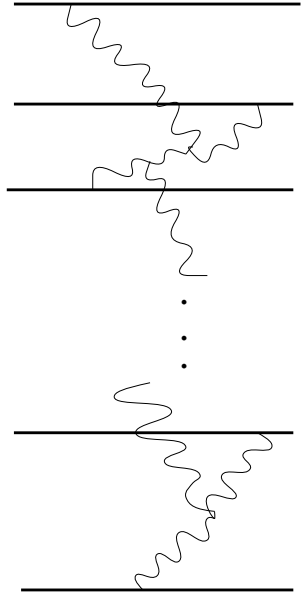,width=4cm,angle=-90}}
\caption{Diagram contributing to $n$ graviton scattering.}
\label{ngraviton}
\end{figure}

\subsection{Other Dimensions}

According to the
Matrix model hypothesis,
the compactification of $M$-theory to $11-k$ dimensions
is described by $k+1$ dimensional super Yang-Mills
theory\cite{compactification}.
For graviton-graviton scattering, this
has been done in \cite{bc}.
It is a simple matter to extend our analysis
to these cases.  

As a simple illustration, consider the three graviton case. 
Working in units where the compact dimensions have $R_k=1$,
then the Fourier transforms needed to convert the supergravity result
(\ref{sugra3}) to an effective potential are
\beq
\frac{\kappa^2 v_3^4}{4(2\pi)^{1+k}} \int \frac{d^{9-k}q_3}{(2\pi)^{9-k}}
\frac{e^{i \vec q_3 \cdot \vec R}}{q_3^2}=\frac{v_3^4}{2^{1+k} 
(\sqrt{\pi})^{1+k}} \frac{1}{R^{7-k}} \Gamma(\frac{7-k}{2})
\eeq
and
\beq
\frac{2\kappa^2 v_{12}^i v_{12}^j} {(2\pi)^{1+k}} 
\int \frac{d^{9-k}q_1}{(2\pi)^{9-k}}
\frac{e^{i \vec q_1 \cdot \vec r}}{q_1^4}=\frac{v_{12}^i v_{12}^j} {2^{k} 
(\sqrt{\pi})^{1+k}} \frac{1}{r^{5-k}} \Gamma(\frac{5-k}{2}).
\eeq
On the matrix
model side, the loop integrals arising from integrating out
the massive states must
now be performed in $k+1$ dimensions giving
\beq
-6 v_3^4 \int \frac{d^{1+k}p}{(2\pi)^{1+k}}
\frac{1}{(p^2 + R^2)^4}
=\frac{v_3^4}{2^{1+k} 
(\sqrt{\pi})^{1+k}} \frac{1}{R^{7-k}} \Gamma(\frac{7-k}{2})
\eeq
and
\beq
4 v_{12}^i v_{12}^j \int \frac{d^{1+k}p}{(2\pi)^{1+k}}
\frac{1}{(p^2 + r^2)^3}
=\frac{v_{12}^i v_{12}^j} {2^{k} 
(\sqrt{\pi})^{1+k}} \frac{1}{r^{5-k}} \Gamma(\frac{5-k}{2})
\eeq
in agreement with the supergravity result above. 
All of the
integrals are convergent for $k \le 4$ and since these same integrals
are needed for our $n$-graviton result, it 
is a simple matter to show that the agreement
we have found here persists for arbitary $n$.

\section{Some Puzzles}

In the original discussion of \cite{bfss}, as well as in \cite{dlcq},
the question was raised:  why does the
lowest order matrix model calculation reproduce the
tree level supergravity result for graviton-graviton
scattering.  The scattering amplitude is given
by a power series in
$g N \over r^3$, and one ultimately
wants to take a limit with $N \rightarrow \infty$,
$g$ fixed.  Moreover, one wants to take this limit
uniformly in $r$, i.e. one does not expect to scale
distances with $N$.  The answer suggested by these authors was
that the explanation lies in a non-renormalization theorem
for $v^4$ terms, which insures that the one-loop result
is exact.  The required cancellation was
demonstrated at two loops in \cite{beckers}.
Such a theorem for four derivative terms in four dimensional field
theory was proven in \cite{ds}.
The complete proof for the matrix model was finally provided in
\cite{sethi}. 

The agreement of three graviton scattering in the matrix
model with supergravity suggests that there are more non-renormalization
theorems governing the various possible terms at order $v^6$.
Indeed, in the case of $SU(2)$, a proof was provided
in \cite{sethitwo}.  On the other hand, it is rather easy to see,
following reasoning similar to that of \cite{echols}, that there
are operators at order $v^6$ which are renormalized in $SU(N)$,
$N \ge 4$.
In particular, consider the case of four gravitons.
In the previous section, we computed the contribution
to the amplitude (\ref{matrix4})
by contracting $x^a x^a$ in eqn.(\ref{prop}), and took
the piece quadratic in $v^2$.  Taking, instead,
the leading, velocity-independent term in this propagator
yields a contribution to the effective action,
\begin{equation}
N_1 N_2 N_3 N_4 {45 \over 256} 
v_4^4 (\vec v_3 \cdot \vec \nabla_4)^2 {1 \over x_4^7}\nabla_3^2
{1 \over x_3^5}
{1 \over x_{12}}.
\label{trouble}
\end{equation}
%
Not only does this represent a renormalization of the
$v^6$ terms computed at two loops, but the $N$-dependence of (\ref{trouble})
is not appropriate to a Lorentz-invariant amplitude.
One might wonder if this term can be cancelled by terms generated
at higher order in the Born series.  However, it is easy
to see that this is not the case.  One can define an index
of an amplitude, ${\cal A}$ (written in momentum
space), $I_{\cal A}$, which is simply the difference
of the number of powers of momentum in the numerator
and in the denominator.
All of the amplitudes we have studied previously
have $I_{\cal A}=2$.
The iterations of the lower order
matrix Hamiltonian also have $I_{\cal A}=2$.
However, (\ref{trouble})
has $I_{\cal A}=-4$.
So this can not be the source of the discrepancy.
We have checked carefully for other diagrams in the
matrix model effective action which might have
this structure, and we do not believe there are any.

So we seem to have found,
consistent with \cite{yoneya}, that the three
graviton amplitude is in agreement with the low energy limit
of supergravity.  We have
also shown such agreement for certain terms of order
$v^{2n}$ in the $n$-graviton scattering amplitude;
further tests of the $v^{2n}$ terms in $n$-graviton
scattering will be reported elsewhere.
But the contribution in (\ref{trouble})
does not seem to correspond to anything
in the DLCQ of supergravity.  
%
If it is correct that the naive DLCQ of supergravity agrees
with the matrix model only in cases where there are non-renormalization
theorems, the work reported here suggests that there are
non-renormalization theorems for the $v^{2n}$ operators
in $SU(n)$, but not for smaller powers of $v$ (in $SU(n)$).
It is perhaps possible to show this using reasoning along
the lines of \cite{sethi,sethitwo}.  We are currently investigating
this possibility.
Our results also seem to be compatible with the results of \cite{ooguri},
who find that two particle scattering is not described correctly
at finite $N$ in curved backgrounds.  In the present case,
one can think of the two ``far away'' gravitons as providing
a background for the gravitons $1$ and $2$.  
The background preserves none of the supersymmetry.
We are currently investigating whether this connection
can be made precise.

\noindent
{\bf Acknowledgements:}

\noindent
We thank T. Banks, G. Ferretti, W. Fischler,  H. Ooguri, J. Plefka,
J. Polchinski,
A. Rajaraman, S. Sethi, S. Shenker, L. Susskind,
W. Taylor, A.  Waldron and A. Wilkins
for discussions.  This work was supported in part by the U.S.
Department of Energy.


\end{document}